\documentclass[pra,showpacs,preprintnumbers,amsmath,amssymb]{revtex4}
\usepackage{graphics}
\usepackage{dcolumn}
\usepackage{bm}

\begin{document}

\title{Negative group delay for Dirac particles travelling
through a potential well}

\author{Xi Chen}
\email{xchen@mail.shu.edu.cn} \affiliation{Department of Physics,
Shanghai University, 99 Shangda Road, Shanghai 200436, P. R. China}%

\author{Chun-Fang Li}
\email{cfli@mail.shu.edu.cn} \affiliation{Department of Physics,
Shanghai University, 99 Shangda Road, Shanghai 200436, P. R.
China} \affiliation{State Key Laboratory of Transient Optics
Technology, Xi'an Institute of Optics and Precision Mechanics,
Academia Sinica, 234 West Youyi Road, Xi'an 710068, P. R. China
}%


\begin{abstract}
The properties of group delay for Dirac particles travelling through a potential well are
investigated. A necessary condition is put forward for the group delay to be negative. It
is shown that this negative group delay is closely related to its anomalous dependence on
the width of the potential well. In order to demonstrate the validity of stationary-phase
approach, numerical simulations are made for Gaussian-shaped temporal wave packets. A
restriction to the potential-well's width is obtained that is necessary for the wave
packet to remain distortionless in the travelling. Numerical comparison shows that the
relativistic group delay is larger than its corresponding non-relativistic one.
\end{abstract}

\maketitle

\section{Introduction}

The question of how much time it takes quantum particles to tunnel through a potential
barrier has been controversial for these decades \cite{Hauge-S,Nimtz,Chiao-S}.
Theoretical investigations \cite{MacColl,Hartman,Buttiker-L,Martin-L,Steinberg-Chiao} and
experimental researches \cite{Steinberg-Kwiat,Carniglia,Enders-1,Enders-2,Sp-R,Balcou}
show that the group delay for some kinds of barriers, also known as the phase time in the
literature, which describes the motion of a wave packet peak \cite{Wigner}, has the
well-known superluminality. In addition, the faster-than-light propagation was also
predicted \cite{Garrett} and experimentally verified \cite{Chu,Segard,Wang-1} for light
pulses through anomalous dispersion media. In a previous paper \cite{Li-Wang}, Li and
Wang  have elaborated the superluminal and even negative properties of the group delay
for quantum particles travelling through a potential well, instead of tunnelling through
a potential barrier. This counterintuitive phenomenon resulted from the interference of
multi-reflected waves in the potential well was demonstrated in a microwave analogy
experiment \cite{Vetter}. Recently, the concept of negative group delay has been extended
to microelectronics \cite{Daniel}.

Most of the theoretical works on tunnelling times in quantum mechanics rely on
Schr\"{o}dinger's non-relativistic theory. Such a theory has a potential deficiency in
accurately addressing the question of causality \cite{Krekora-1,Li-Chen}. For this
reason, a few authors \cite{Krekora-1,Li-Chen,Leavens,Petrillo} extended the analysis of
the tunnelling to Dirac's fully relativistic quantum theory. Leavens and Aers
\cite{Leavens} used the stationary-state method to analyze the Larmor-clock transmission
times for single barriers and resonant double-barriers. Krekora {\it et al.}
\cite{Krekora-1} solved numerically the time-dependent Dirac equation for a quantum wave
packet tunnelling through a potential barrier. And Petrillo and Janner \cite{Petrillo}
studied the dynamics of wave-packet tunnelling through a barrier. In a recent work
\cite{Li-Chen}, we discussed an energy-transfer associated traversal time for Dirac
particles tunnelling through a potential barrier.

The purpose of this paper is to investigate the properties of the group delay for Dirac
particles travelling through a potential well. It is shown that it behaves superluminal
and even negative in much the same way as in the non-relativistic quantum mechanics
\cite{Li-Wang}. The negativity of the group delay is closely related to its anomalous
dependence on the width of the potential well around transmission resonances. In order to
demonstrate the validity of the stationary-phase approximation, numerical simulations are
made for Gaussian-shaped temporal wave packets. A restriction to the width of the
potential well is given that is necessary for the wave packet to remain distortionless in
the travelling. Finally, a numerical comparison shows that the Dirac's relativistic group
delay is larger than its corresponding non-relativistic one.

\section{Relativistic group delay and its non-relativistic limit}

Consider Dirac particles of precisely defined incident energy $E$
and of helicity $+1$, travelling through a one-dimensional
rectangular potential well $-V_0\Theta(z)\Theta(a-z)$ (with $V_0$
positive, representing the depth of the potential well). Let the
incident wave function be
\begin{equation} \label{incident wave}
\psi_{in}(z)=\left(\begin{array}{c} 1 \\ 0 \\ \frac{\hbar kc}{E +
\mu c^2} \\ 0
\end{array}\right) e^{ikz},(z<0),
\end{equation}
where $k=(E^2-\mu^2c^4)^{1/2}/\hbar c$, $\mu$ is the mass of incident particles, and $c$
is the speed of light in vacuum. Then Dirac equation and boundary conditions give for the
transmitted wave function
\begin{equation}
\label{transmitted wave} \psi_{tr}(z)=F\left(\begin{array}{c} 1 \\ 0 \\
\frac{\hbar kc}{E+\mu c^2} \\ 0
\end{array}\right) e^{ik(z-a)},(z>a),
\end{equation}
where the transmission coefficient $F=e^{i\phi}/f$ is determined by the following complex
number,
$$
fe^{i \phi}=\cos{k'a}+(i/2)(\chi+1/\chi) \sin{k'a},
$$
so that
\begin{equation}
\label{phase} \phi=
\mbox{int}\left(\frac{k'a}{\pi}+\frac{1}{2}\right)\pi+
\tan^{-1}\left[\frac{1}{2}\left(\chi+\frac{1}{\chi}\right)\tan{k'a}\right],
\end{equation}
int($\cdot$) stands for the integer part of involved number,
$k'=[(E+V_0)^2-\mu^2c^4]^{1/2}/\hbar c$. Noting that the real
parameter $\chi$ is defined as
$$\chi \equiv \frac{k}{k'}\frac{E+V_0+\mu
c^2}{E+\mu c^2},$$ and has the property that $0<\chi<1$.
Obviously, $\phi$ is the phase shift of transmitted wave
(\ref{transmitted wave}) at $z=a$ with respect to the incident
wave (\ref{incident wave}) at $z=0$. The transmission probability
$T$ is a periodical function of the width $a$ of the potential
well,
\begin{equation}
\label{transmission probability} T=\frac{1}{f^2}=\frac{4 \chi^2}{4
\chi^2+(\chi^2-1)^2 \sin^{2} {k' a}}.
\end{equation}

The group delay, defined as the derivative of the phase shift $\phi$ with respect to
particle's energy $E$ \cite{Chiao-S,Wigner}, is given by
\begin{eqnarray}
\label{group delay-1} \tau_{\phi}&=& \hbar \frac{\partial
\phi}{\partial E}=\frac{T}{2 \chi \hbar k'
c}\left[(1+\chi^2)(E+V_{0}) -(1-\chi^2) \frac{\mu
V_{0}(2E+V_{0})}{\hbar^2 k^2} \frac{\sin {2k'a}} {2k'a}\right]
\frac{a}{c}.
\end{eqnarray}
For the following comparisons, let us first give its non-relativistic limit. By
non-relativistic limit it is meant that the speed of incident particles is much smaller
than $c$, so that their kinetic energy $E'=E-\mu c^2$ is much smaller than their rest
energy $\mu c^2$, $E' \ll \mu c^2$. It is also meant that the interaction energy $V_0$
satisfies $V_0 \ll \mu c^2$. In this limit, we have $k \approx \sqrt{2 \mu E'}/\hbar$,
$k' \approx [2 \mu (E'+V_0)]^{1/2}/\hbar$, and $\chi \approx k/k'$. Collecting all these,
we get
\begin{equation}
\label{group delay-2} \tau_{\phi} \approx \tau'_{\phi}=\frac{2 \mu
a}{\hbar
k}\frac{k^2(k^2+k'^2)/k_0^4-\sin{2k'a}/{2k'a}}{4k^2k'^2/k_0^4+\sin^2{k'a}},
\end{equation}
which is expected from Schr\"{o}dinger's non-relativistic theory
\cite{Li-Wang}, where $k_0 \approx \sqrt{2 \mu V_0}/\hbar$.

\section{Negative property of the group delay}

In this section, we discuss the negative property of the group
delay. It is seen from Eq. (\ref{group delay-1}) that when
inequality
$$
(1+\chi^2)(E+V_0)<(1-\chi^2)\frac{\mu
V_0(2E+V_0)}{\hbar^2k^2}\frac{\sin2k'a}{2k'a}
$$
holds, the group delay is negative, $\tau_\phi<0$. Since
$\sin2k'a/2k'a<1$, the above inequality leads to the following
necessary condition for the group delay to be negative,
\begin{equation} \label{necessary condition}
(1+\chi^2)(E+V_0) < (1-\chi^2) \frac{\mu V_0(2E+V_0)}{\hbar^2k^2},
\end{equation}
which can be expressed as a restriction to the total energy of
incident particles as follows,
\begin{eqnarray}  \label{inequality-1}
E<E_t &\equiv& \mu c^2 \left\{\frac{V_0}{2\mu c^2}+\left[\left(\frac{V_0}{2\mu
c^2}\right)^2-\left(\frac{1}{3}\right)^3\right]^{1/2}\right\}^{1/3} \nonumber \\ && + \mu
c^2 \left\{\frac{V_0}{2\mu c^2}-\left[\left(\frac{V_0}{2\mu
c^2}\right)^2-\left(\frac{1}{3}\right)^3\right]^{1/2}\right\}^{1/3}.
\end{eqnarray}
This means that when the energy of incident particles satisfies
Eq. (\ref{inequality-1}), that is to say, the energy of incident
particles $E$ is less than a threshold energy $E_t$, one can
always find a width $a$ of the potential well at which the group
delay is negative. Of course, $E_t$ in Eq. (\ref{inequality-1}) is
always real and larger than $\mu c^2$. In the case of $V_0 < 2\mu
c^2/3\sqrt{3}$, $E_t$ can be rewritten as
\begin{equation}
\label{E-t} E_t = \mu c^2 \left\{\frac{V_0}{2\mu c^2}+i
\left[{\left(\frac{1}{3}\right)}^3-\left({\frac{V_0}{2\mu
c^2}}\right)^2\right]^{1/2}\right\}^{1/3} + \mbox{c.c}.
\end{equation}
In the non-relativistic limit, $V_0 \ll \mu c^2$, we obtain $E_t \approx \mu c^2+ V_0/2$.
Therefore, the necessary condition (\ref{inequality-1}) now reduces to $E'<V_0/2$, as
observed previously in Schr\"{o}dinger's theory \cite{Li-Wang}.

\begin{figure}[ht]
\includegraphics{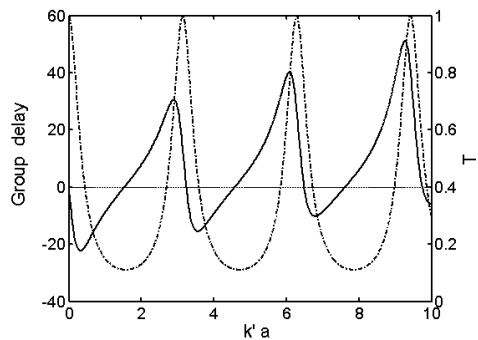}
\caption{\label{fig.1} Dependence of the group delay (in the unit of $\hbar/\mu c^2$) on
potential-well's width $a$, where $E=1.01\mu c^2$, $V_0=0.4\mu c^2$, and $a$ is re-scaled
to be $k'a$. Here the periodical dependence of transmission probability $T$ on $a$ is
also depicted by dashed curve.}
\end{figure}

Fig. \ref{fig.1} shows a typical example of the dependence of $\tau_\phi$ on the width
$a$, where the well depth $V_0=0.4 \mu c^2 > 2 \mu c^2/3 \sqrt{3}$ ($E_t=1.16 \mu c^2$),
the total energy $E=1.01 \mu c^2 < E_t$, and $a$ is re-scaled to be $k'a$. For
comparison, Fig. \ref{fig.1} also shows the periodical dependence of transmission
probability $T$ on $a$ under the same conditions. It is interesting to note that the
oscillation of the group delay with respect to $a$ is closely related to the periodical
occurrence of transmission resonances at $k'a=m \pi$ ($m=1,2,3...$).

On the one hand, at resonances, the group delay becomes
\begin{equation}
\tau_\phi|_{k'a=m\pi}=\frac{1}{2}\left(\chi+\frac{1}{\chi}\right)
\frac{E+V_0}{ \hbar k'c}\frac{a}{c}>\frac{a}{c},
\end{equation}
which is proportional to $a$. Its corresponding group velocity is less than the velocity
of light in vacuum, $c$. On the other hand, the derivative of group delay $\tau_\phi$
with respect to $a$ is, at resonances,
\begin{eqnarray}
\frac{\partial \tau_\phi}{\partial a}|_{k'a=m\pi}&=& \frac{1}{2
\chi \hbar k'c^2}\left[(1+\chi^2)(E+V_0)
-(1-\chi^2)\frac{\mu V_0(2E+V_0)}{%
\hbar^2 k^2}\right].
\end{eqnarray}
When the necessary condition (\ref{necessary condition}) is satisfied, it is negative.
This shows that the group delay depends anomalously on the width around resonance points.
In other words, it decreases with increasing the width of the potential well at
resonances, as displayed clearly in Fig. \ref{fig.1}.

\begin{figure}[ht]
\includegraphics{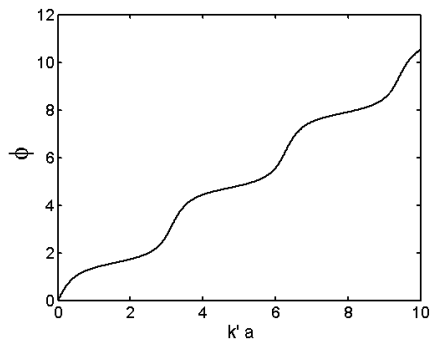}
\caption{\label{fig.2} The quantum-like behavior of phase shift (\ref{phase}) with
respect to the width of potential well, $a$, where $E=1.01 \mu c^2$ and $V_0=0.4 \mu
c^2$}
\end{figure}

In connection with the negative characteristics of the group
delay, the phase shift (\ref{phase}) shows a quantum-like behavior
with respect to the potential-well's width.
Fig. \ref{fig.2} indicates such a behavior, where $%
E=1.01\mu c^2$ and $V_0=0.4\mu c^2$. This quantum-like behavior is
also related to the periodical occurrence of the transmission
resonances. It can be seen from Eq. (\ref{phase}) and Fig.
\ref{fig.2} that at resonances, $k'a=m\pi$, the phase shift
becomes $\phi=k'a$ and changes rapidly around here with respect to
$a$. However, when the width is far from the resonance points as
is in the middle between two adjacent resonance points, the phase
shift changes slowly.

\begin{figure}[ht]
\includegraphics{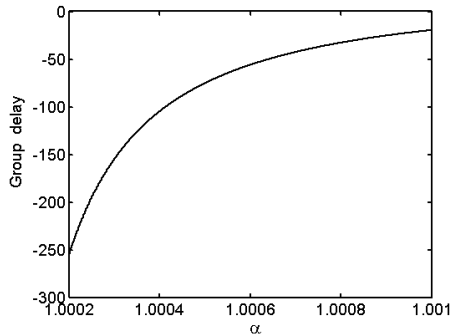}
\caption{\label{fig.3} Dependence of the group delay (in the unit
of $\hbar/\mu c^2$) on the incident energy $\alpha$, where
$\beta=0.4$ and $\gamma=0.01$.}
\end{figure}

In addition, it is also indicated from Eq. (\ref{group delay-1})
that the group delay depends not only on the potential-well's
width $a$, but also on the incident energy $E$ and the depth of
the potential well $V_0$. To see the latter more clearly, we
convert it to a dimensionless form. Denoting $\alpha=E/\mu c^2$,
$\beta=V_0/\mu c^2$, and $ \gamma=\hbar/a\mu c$, we have
\begin{eqnarray} \label{dimensionless}
\frac{\tau_\phi}{\tau_0} &=& \frac{T}{2\chi} \frac{k'a}
{(\alpha+\beta)^2-1} \left[(1+\chi^2)( \alpha+\beta) - (1-\chi^2)
\frac{\beta(2\alpha+\beta)}{\alpha^2-1}\frac{\sin
{2k'a}}{2k'a}\right],
\end{eqnarray}
where $\tau_0=\hbar/\mu c^2$, $k'a=[(\alpha+\beta)^2-1]^{1/2}/ \gamma$. When the kinetic
energy of incident particles $E'$ is small enough that $E'/\mu c^2 \rightarrow 0$ with
the depth of the potential well remaining finite, we have $\alpha \rightarrow 1$, so that
$\chi \rightarrow 0$. In this limit, the group delay (\ref{dimensionless}) has the
following form,
$$
\lim_{\alpha\rightarrow 1}\frac{\tau_\phi}{\tau_0}= -\sqrt{\frac{
\beta+2}{(\alpha^2-1)\beta}}\cot{k'a},
$$
which approaches negative infinite when $\cot{k'a}>0$. Fig. \ref{fig.3} shows such a
dependence of the group delay on the incident energy $E$, where $\beta=0.4$ and
$\gamma=0.01$. A strange phenomenon occurs here that for a given potential well, the
absolute value of the negative group delay becomes larger when decreasing the incident
kinetic energy. Of course, the transmission probability in this limit tends to zero in
the following way,
\begin{equation}  \label{transmission limit-1}
\lim_{\alpha \rightarrow 1}T=\frac{4\chi^2}{4\chi^2+\sin^2{k'a}},
\end{equation}
so that very few particles can travel through the potential well
at this negative group velocity.

In order to demonstrate the validity of the above stationary-phase method in this
problem, we proceed to numerical simulations of the group delay for a Gaussian-shaped
wave packet. The incident wave function of Dirac particles is assumed to be, at $z=0$,
\begin{equation}
\label{wavepacket} \Psi_{in}(t)|_{z=0}=\left(\begin{array}{c} 1 \\ 0 \\
\frac{\hbar k_0c}{E_0 + \mu c^2} \\ 0
\end{array}\right) \exp(-t^2/2 w^2 -iE_0 t /\hbar),
\end{equation}
which has the Fourier integral of the following form,
\begin{equation} \label{incidentintegral}
\Psi_{in}(t)|_{z=0}=\frac{1}{\sqrt{2\pi}}\int
A(E)\psi(E_0)\exp(-iE t/\hbar)dE,
\end{equation}
where $k_0=(E_0^2-\mu^2 c^4)^{1/2}/\hbar c$, the spinor $\psi(E_0)$ is $[1 , 0, \hbar
k_0c/(E_0 + \mu c^2), 0 ]^T$, the energy spectral distribution $A(E)$ with the central
energy $E_0$ is given by
$$A(E)=(w/\hbar)\exp[-(w^2 /2\hbar^2)(E-E_0)^2],$$
and $w$ is the temporal width of the wave packet (\ref{wavepacket}). The transmitted wave
function takes the following form,
\begin{equation}
\label{transmittedintegral} \Psi_{tr}(z,t)=\frac{1}{\sqrt{2\pi}}\int
F(E)A(E)\psi(E_0)\exp\{i[k(z-a)-Et/\hbar]\}dE,
\end{equation}
The numerically calculated group delay, $\tau^N _{\phi}$, is defined here by
\begin{equation}
\label{numerical result} |\Psi_{tr}(z=a,\tau^N _{\phi})|^2=\mbox{max}\{|\Psi_{tr}
(z=a,t)|^2 \}.
\end{equation}
Since the incident wave is of perfect Gaussian shape, the integral limit in Eq.
(\ref{incidentintegral}) and hence in Eq. (\ref{transmittedintegral}) should be from
$-\infty$ to $+\infty$. But the energy of incident particles must be larger than $\mu
c^2$. So the real integral in numerical simulations is taken to be from $\mu c^2$ to
$+\infty$.

\begin{figure}[ht]
\includegraphics{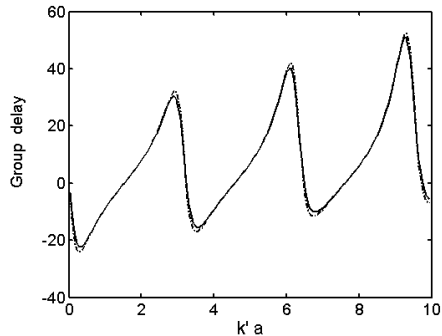}
\caption{\label{fig.4} Comparison of theoretical  and  numerical
results of group delays (in the unit of $\hbar/\mu c^2$) with
respect to $a$, where $E_0=1.01 \mu c^2$, $V_0=0.4 \mu c^2 $,
$w=300 \tau_0$ and $a$ is re-scaled to $k'a$. The theoretical
result for the group delay (\ref{group delay-1}) is depicted by
the real curve, and the dash curve corresponds to the numerical
result for the group delay defined by ($\ref{numerical result}$).
}
\end{figure}

Calculations show that the stationary-phase approximation (\ref{group delay-1}) for the
group delay is in good agreement with the numerical result, especially when the energy
spectral distribution is sharp. In Fig. \ref{fig.4} we show such a comparison between
theoretical and numerical results, where $E_0=1.01 \mu c^2$, $V_0=0.4 \mu c^2$, and the
temporal width $w=300 \tau_0$. For the chosen temporal width, the corresponding energy
spreading $\Delta E=\hbar/2w=\mu c^2/600$ is narrow enough that the integral in Eq.
(\ref{transmittedintegral}) can be performed from $\mu c^2$ to $2 \mu c^2$ without
changing the result significantly.

As pointed out in the optical analog \cite{Huang}, for an incident wave packet of energy
spreading $\Delta E$, the corresponding spreading of $k'a$ should be much smaller than
$\pi$, the period of $|F|$, in order that the stationary-phase approximation is valid.
With the energy spreading $\Delta E=\hbar/2w$, this leads to the following restriction to
the width of the potential well,
\begin{equation}
a\ll 2 \pi w \frac{\partial E}{\hbar \partial k'}.
\end{equation}
It is noted that $\partial E/\hbar \partial k'$ is nothing but the group velocity of
particles in the region of potential well. By introducing a characteristic length $L$
which is defined as $w \partial E/\hbar \partial k'$, the above restriction is simplified
to be $a \ll 2\pi L$. With this restriction, the temporal wave packet can travel through
the potential well with negligible distortion.

All the above discussions show that the group delay (\ref{group delay-1}) in Dirac's
relativistic theory can be superluminal and even negative in much the same way as in
Schr\"{o}dinger's non-relativistic theory \cite{Li-Wang}. In this sense, the superluminal
and negative properties is not an artifact due to the deficiency of non-relativistic
quantum theory. Rather, it is a real effect. In the next section, we investigate the
difference between the group delays in relativistic and non-relativistic theories.

\section{Relativistic effect on the group delay}

As mentioned in Sec.II, when the incident kinetic energy $E' \ll \mu c^2$ and the
potential-well's depth $V_0 \ll \mu c^2$, the group delay (\ref {group delay-1}) tends to
its non-relativistic limit (\ref{group delay-2}). To show their differences, we rewrite
the non-relativistic group delay in the following dimensionless form,
\begin{eqnarray}  \label{dimensionless2}
\frac{\tau'_\phi}{\tau_0} &=& \frac{k'a} {\sqrt{
(\alpha-1)(\alpha+\beta-1)}}\frac{(\alpha -1)(2\alpha+\beta-2)
-\beta^2\sin{2k'a}/{2k'a}} {4(\alpha-1) (\alpha+\beta-1)+ \beta^2
\sin^2{k'a}},
\end{eqnarray}
where $\alpha$, $\beta$, and $\gamma$ are the same as before, $k'a
= [2(\alpha+\beta-1)]^{1/2}/\gamma$ in the non-relativistic
quantum mechanics.

\begin{figure}[ht]
\includegraphics{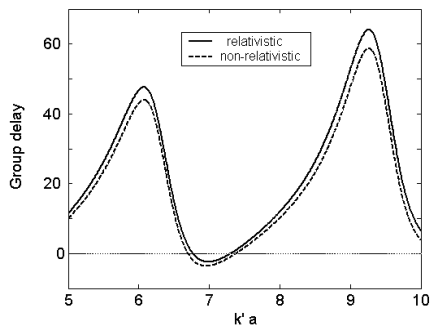}
\caption{\label{fig.5} Comparison of relativistic and
non-relativistic group delays (in the unit of $\hbar/\mu c^2$) for
$\alpha=1.01$ and $\beta=0.2$. The real curve denotes the
relativistic group delay (\ref{dimensionless}), and the dashed
curve corresponds to the non-relativistic group delay
(\ref{dimensionless2}). }
\end{figure}

In order to address the relativistic effect on the group delay, we draw in Fig.
\ref{fig.5} the dependence of relativistic and non-relativistic group delays on the width
of the potential well for $\alpha=1.01$ and $\beta=0.2$ ($V_0=0.2 \mu c^2<2 \mu c^2
/3\sqrt{2}$), where the relativistic group delay is shown by solid curve, the
corresponding non-relativistic one is shown by dashed curve, and the width of potential
well is re-scaled by their own $k'$ to be $k'a$.

From Fig. \ref{fig.5} we see that the relativistic group delay for travelling through a
potential well is larger than the corresponding non-relativistic one. This is in
agreement with the result of Leavens and Aers \cite{Leavens}, who observed that the local
mean velocity for transmitted particles is reduced due to relativistic effect for
resonant double barriers where the group delay is always larger than zero. The difference
between the two group delays are very small for the low energy and potential-well depth.
It is expected that the agreement between the group delays in relativistic and
non-relativistic theory is obtained when $E' \ll \mu c^2$ and $V_0 \ll \mu c^2$.

\section{Conclusions}

In summary, we have investigated the negative property of the group delay for Dirac
particles travelling through a quantum potential well. A necessary condition
(\ref{necessary condition}) is given for the group delay to be negative, which is a
restriction on the energy of incident particles and reduces to $E'<V_0/2$ in
non-relativistic limit. The relation of the negativity of the group delay with its
anomalous dependence on the width of the potential well around transmission resonances is
discussed. In order to demonstrate the validity of the stationary-phase approach,
numerical simulations are made for a Gaussian-shaped temporal wave packet. A restriction
to the width of the potential well is given that is necessary for the wave packet to
remain distortionless in the travelling. Comparison of the relativistic and
non-relativistic group delays is also made. It is found that the value of the
relativistic group delay is larger than that of the non-relativistic one. It should be
pointed out that the negative group delay discussed here is not at odds with the
principle of causality. When a wave packet travels through a potential well, the boundary
effect leads to a reshaping of the edge of the packet \cite{Steinberg-Kwiat,Japha} which
results in an effective acceleration of the wave packet. This phenomenon is similar to
but different from the negative group-velocity propagation performed by Wang {\it et.
al.} \cite{Wang-1} where the gain-assisted anomalous dispersion plays the role. We hope
that this work may stimulate further experimental researches in electronic domains.

\section*{Acknowledgments}

This work was supported in part by the Science Foundation of Shanghai Municipal
Commission of Education (Grant No. 01SG46), the Science Foundation of Shanghai Municipal
Commission of Science and Technology (Grant No. 03QMH1405), and by Shanghai Leading
Academic Discipline Program.

\end{document}